# Using High Frequency Propagation to Calculate Basic Maximum Usable Frequency

Israa Abdualqassim Mohammed Ali[*]

[*] University of Baghdad, College of Science, Baghdad, Iraq

**ABSTRACT**

A comparison between observed (obs.) digital ionospheric sounding data and predicted (pre.) using International Reference Ionosphere (IRI) model for critical frequency (foF2) and Basic Maximum Usable Frequency (BMUF) of ionospheric F2-Layer has been made. A mid-latitude region selected for this research work by using data from station Wakkanai (45.38$^o$ N, 141.66$^o$ E). This study included 12 monthly median data from year (2001, $R_{12}$=111) selected for high Sunspot number (SSN) and years for low SSN (2004, $R_{12}$=44) and (2005, $R_{12}$=29). Frequency parameters foF2 reveals that there is a good correlation between observed and predicted except for January of years 2001 and 2004, and BMUF revealed that there is a good correlation between observed and predicted for years of low SSN and all months except in month 1, 9 and 12 of year 2004, for year 2001 of high SSN there is a bad correlation. A correction factor as a function of time used from fitting technique to correct the predicted value with observed value of BMUF for year 2001.

Author Correspondence, e-mail: israa.aq88@gmail.com

## 1. INTRODUCTION

The maximum usable frequency (MUF) is important ionospheric parameter for radio users because of its role in radio frequency management and for providing a good communication link between two locations (Athieno et al., 2015; Suparta et al., 2018). MUF would be used to designate the highest signal frequency for high frequency (HF) communications that can be used for radio transmission between two points by reflection from the ionosphere at a given time under specific ionospheric conditions (Harris, 2005; Freeman, 2006; Mudzingwa et al., 2013) which fluctuate continuously because the ionosphere acts as a dispersive medium (Athieno et al., 2015). MUF is simply given by (Fotiadis et al., 2004; Oyekola, 2010; Malik et al., 2016):

MUF = $fo$F2 ✕ M(3000)F2,

where foF2 is the critical frequency of the F2 layer, i.e., the highest frequency that would be reflected by the ionosphere at vertical incidence (McNamara et al., 2007; Harish et al., 2009), and M (3000) F2 is the propagation factor of layer F2 in which represents the optimal frequency to broadcast a signal received at a distance of 3000 km (Kouris et al., 2000; Adeniyi et al., 2003; Oyekola et al., 2012; Nagar et al., 2015).

The main drawback of the HF ionospheric communications is that the characteristics of the transmission medium strongly influenced by space weather that is varying over time, such as the Sunspot Number (SSN). Ionospheric effects of space weather would give rise to sudden changes in the spatial structure and density of ionospheric regions. These variations then affect HF radiowave propagation by altering parameters such as the MUF over HF communications links (Hughes et al., 2002; Bahari et al., 2018). Since HF communications has wide range of application areas, the measurements with the real-time of HF propagation conditions is a key factor of space weather monitoring systems to get satisfactory performance.

The aim of the present research work is to study a comparison between observed digital ionospheric sounding data and predicted using IRI model for (foF2) and (BMUF) of ionospheric F2-Layer of a mid-latitude region.

## 2. DATA SELECTION

Ionosonde data station was used for comparison with IRI model, the station chosen was Wakkanai station (latitude 45.38oN, longitude 141.66oE) for calculating basic monthly median MUF from observed data of the F2 layer critical frequency (f0F2) and M3000F2 for selected years (2001, 2004 and 2005). The degree of correlation between daily sunspot number and daily values of ionospheric paramters is not as high as in the case of the monthly median values taken from Sunspot Bulletin as mentioned by Kouris et al. (1998). Therefore, daily sunspot number is not a good indicator and monthly median sunspot data used instead of day-to-day predictions. Sunspot number (SSN) selected from Solar Influences Data analysis Center (SIDC). Table 1 reveals the average 12 month SSN for all years selected.

## 3. DATA ANALYSES

To study the BMUF we need to calculate the critical frequency (predicted) foF2 and M3000F2 from the IRI model by using CCIR coefficient for years selected (2001) for high SSN and (2004, and 2005) for low SSN and compared with data observed. In figure 6 is a good correlation between observed BMUF and predicted for years (2004,2005) for SSN low but for years (2001) for high SSN, from figure 6 there is a bad correlation between observation and predicted BMUF, it can be corrected by fitting taken correction formula (eq. 1) for 24 hour. Table 2 represents the fitting coefficients $a_0$, $a_1$, and $a_2$. Tables 3 and 4 reveal the absolute error between observation and prediction values before and after correction respectively for 24 hours and 12 months for year 2001.

$$\text{Corr} = a_0 + a_1 T + a_2 T^2 \tag{1}$$

$$\text{Absolute error} = \sqrt{(\text{Obs.} - \text{Pre.})^2} \tag{2}$$

## 4. RESULTS AND DISCUSSION

Figures (1 and 2) represent the observed and predicted monthly median foF2 with local time for three years 2001, 2004, and 2005 respectively, in which it reveal that the critical frequency have a high value in the day and low in the night. In winter season foF2 values higher than summer season for high and low SSN this is anomaly in mid-latitude region.

Figures (3, 4, and 5) represent comparison between observed and predicted monthly median

foF2 with local time for three years 2001, 2004, and 2005 respectively. From these figures there is a good correlation between observed and predicted value for all 12 months and year chosen.

Figures (6, 7, and 8) represent comparison between observed and predicted BMUF for years (2001, 2004, and 2005) respectively. From figures it reveal that there is a good correlation between observed and predicted value for all 12 months when SSN low, but there is a bad correlation between observed and predicted value for all 12 months for year 2001 high SSN.

Figure (9) represent the corrected predicted value for year 2001, which is revealed that there is a good correlation between observed and predicted BMUF all 12 month and for 24 hours of the day and night.

## 5. SUMMARY

Space weather effects can seriously impact HF communications by changing the ionospheric environment through which the radio waves propagate. Ionospheric data from the ionosonde stations such as Wakkanai station can be used to monitor this environment by observing Mid-latitude propagation conditions. From drawing foF2 and BMUF with local time for years (2001, 2004, and 2005) it concluded that the observed and predicted monthly median foF2 and BMUF with local time in the day is high than night. Frequency parameters foF2 reveals that there is a good correlation between observed and predicted. BMUF reveal that there is a good correlation between observed and predicted for years of low SSN and all months except in month 1, 9 and 12 of year 2004, for year 2001 of high SSN there is a bad correlation. The predicted value of BMUF for year 2001 can be corrected using correction equation as shown in this study.

## 6. ACKNOWLEDGEMENTS

Table 1. The value of R12 for years 2001, 2004, and 2005

| Month\Year | 2001 | 2004 | 2005 |
|---|---|---|---|
| 1 | 95.6 | 37.3 | 31.3 |
| 2 | 80.6 | 45.8 | 29.2 |
| 3 | 113.5 | 49.1 | 24.5 |
| 4 | 107.7 | 39.3 | 24.2 |
| 5 | 96.6 | 41.5 | 42.7 |
| 6 | 134 | 43.2 | 39.3 |
| 7 | 81.8 | 50.1 | 40.1 |
| 8 | 106.4 | 40.9 | 36.4 |
| 9 | 150.7 | 27.7 | 21.9 |
| 10 | 125.5 | 48 | 8.7 |
| 11 | 106.5 | 43.5 | 18 |
| 12 | 132.2 | 17.9 | 41.1 |
| Ave. | 111 | 44 | 29 |

Table 2. Coefficients $a_o$, $a_1$, and $a_2$ of corrected predict BMUF for year 2001.

| Month\ Coff. | $a_o$ | $a_1$ | $a_2$ |
|---|---|---|---|
| 1 | -0.006 | 0.144 | 0.875 |
| 2 | -0.003 | 0.097 | 1.185 |
| 3 | -2E-05 | 0.012 | 1.822 |
| 4 | -0.000 | 0.000 | 1.789 |
| 5 | 0.000 | - 0.031 | 1.642 |
| 6 | 0.000 | - 0.017 | 1.600 |
| 7 | 0.001 | - 0.046 | 1.812 |
| 8 | 0.000 | - 0.032 | 1.873 |
| 9 | -0.001 | 0.029 | 1.834 |
| 10 | -0.003 | 0.084 | 1.310 |
| 11 | -0.005 | 0.141 | 1.056 |
| 12 | -0.008 | 0.230 | 0.817 |

Table 3. The absolute error between observed and predicted BMUF for year (2001)

| H\M | 1 | 2 | 3 | 4 | 5 | 6 | 7 | 8 | 9 | 10 | 11 | 12 |
|---|---|---|---|---|---|---|---|---|---|---|---|---|
| 0 | 0.8 | 2.5 | 7.3 | 7.9 | 6.4 | 6.9 | 7.0 | 7.1 | 7.7 | 4.2 | 2.6 | 2.3 |
| 1 | 0.7 | 2.5 | 7.2 | 8.3 | 7.8 | 7.2 | 7.3 | 8.0 | 7.4 | 4.3 | 2.8 | 2.2 |
| 2 | 1.0 | 2.9 | 7.2 | 8.0 | 7.5 | 7.3 | 8.3 | 8.5 | 7.9 | 4.9 | 2.8 | 1.8 |
| 3 | 1.6 | 3.3 | 6.6 | 7.5 | 7.8 | 7.1 | 8.3 | 8.0 | 8.0 | 4.5 | 2.8 | 2.3 |
| 4 | 1.6 | 3.6 | 6.3 | 6.2 | 6.7 | 8.0 | 8.5 | 7.4 | 7.5 | 3.9 | 2.5 | 2.7 |
| 5 | 1.9 | 3.9 | 6.0 | 7.8 | 7.2 | 8.3 | 7.7 | 7.6 | 8.5 | 4.8 | 3.9 | 3.2 |
| 6 | 1.9 | 4.2 | 12.6 | 10.5 | 5.7 | 8.2 | 7.8 | 8.4 | 14.5 | 10.3 | 5.2 | 2.2 |
| 7 | 6.7 | 12.2 | 17.2 | 10.7 | 4.2 | 7.4 | 6.8 | 8.9 | 15.6 | 16.1 | 15.5 | 9.5 |
| 8 | 14.2 | 15.9 | 17.7 | 11.5 | 4.1 | 5.9 | 7.8 | 6.7 | 15.2 | 19.7 | 21.0 | 19.6 |
| 9 | 15.9 | 15.8 | 17.0 | 12.1 | 4.7 | 4.2 | 5.7 | 6.8 | 15.0 | 17.1 | 22.0 | 26.8 |
| 10 | 16.2 | 16.5 | 18.7 | 13.4 | 4.6 | 5.0 | 5.5 | 6.7 | 14.3 | 17.1 | 20.8 | 28.5 |
| 11 | 13.7 | 14.6 | 15.4 | 13.4 | 4.5 | 5.4 | 4.2 | 7.0 | 14.0 | 16.2 | 18.6 | 22.9 |
| 12 | 11.8 | 13.6 | 17.4 | 12.3 | 6.4 | 4.7 | 5.5 | 9.5 | 13.7 | 15.3 | 16.0 | 18.7 |
| 13 | 12.3 | 12.7 | 15.1 | 11.2 | 8.8 | 7.4 | 6.5 | 9.6 | 11.7 | 14.3 | 15.0 | 19.8 |
| 14 | 13.0 | 13.7 | 15.4 | 12.9 | 8.3 | 8.4 | 6.6 | 8.9 | 13.1 | 15.5 | 19.1 | 20.2 |
| 15 | 14.1 | 15.6 | 15.2 | 12.1 | 7.4 | 6.8 | 6.8 | 8.8 | 13.0 | 15.6 | 19.1 | 19.6 |
| 16 | 12.0 | 13.9 | 16.2 | 12.5 | 6.3 | 6.4 | 5.2 | 8.2 | 12.3 | 14.7 | 16.9 | 18.7 |
| 17 | 7.1 | 12.2 | 16.8 | 11.9 | 6.1 | 6.1 | 5.7 | 7.1 | 12.4 | 13.9 | 13.1 | 17.7 |
| 18 | 6.4 | 9.7 | 15.3 | 11.4 | 6.2 | 5.4 | 5.8 | 5.7 | 11.6 | 11.5 | 12.1 | 13.7 |
| 19 | 4.3 | 9.0 | 12.3 | 8.9 | 6.4 | 5.3 | 5.5 | 6.6 | 9.1 | 11.2 | 9.7 | 9.6 |
| 20 | 1.5 | 6.1 | 11.9 | 6.3 | 5.5 | 5.0 | 4.6 | 6.0 | 9.2 | 8.8 | 6.7 | 6.0 |
| 21 | 0.7 | 4.2 | 11.0 | 6.5 | 4.6 | 4.3 | 5.4 | 6.1 | 8.2 | 6.5 | 5.2 | 5.0 |
| 22 | 1.6 | 6.0 | 11.3 | 7.1 | 4.6 | 5.8 | 5.9 | 5.5 | 7.7 | 4.8 | 4.3 | 5.3 |
| 23 | 0.9 | 4.3 | 9.8 | 8.5 | 5.3 | 6.7 | 7.1 | 5.9 | 7.8 | 4.2 | 3.2 | 2.4 |
| Ave. | 6.7 | 9.1 | 12.8 | 10.0 | 6.1 | 6.4 | 6.5 | 7.5 | 11.1 | 10.8 | 10.9 | 11.7 |

Table 4. The absolute error between observed and predicted corrected BMUF for year (2001)

| H\M | 1 | 2 | 3 | 4 | 5 | 6 | 7 | 8 | 9 | 10 | 11 | 12 |
|---|---|---|---|---|---|---|---|---|---|---|---|---|
| 0 | 1.98 | 0.70 | 0.62 | 0.25 | 2.03 | 1.64 | 2.98 | 2.52 | 0.95 | 0.66 | 2.04 | 3.81 |
| 1 | 0.56 | 0.18 | 0.75 | 0.72 | 0.74 | 0.56 | 1.24 | 0.18 | 0.89 | 0.02 | 0.85 | 1.85 |
| 2 | 0.26 | 0.36 | 0.45 | 0.59 | 1.34 | 0.29 | 1.06 | 1.75 | 0.27 | 0.07 | 0.17 | 0.19 |
| 3 | 0.66 | 0.58 | 0.89 | 0.12 | 2.13 | 0.69 | 1.98 | 1.71 | 0.09 | 0.81 | 1.21 | 1.29 |
| 4 | 1.57 | 0.91 | 1.29 | 1.40 | 0.92 | 1.83 | 2.40 | 0.93 | 0.28 | 1.90 | 2.38 | 2.02 |
| 5 | 2.01 | 0.98 | 1.53 | 0.68 | 0.81 | 1.92 | 1.00 | 0.29 | 0.17 | 1.65 | 1.67 | 2.35 |
| 6 | 2.87 | 1.53 | 4.47 | 0.33 | 1.55 | 1.06 | 0.33 | 0.36 | 4.31 | 2.18 | 2.08 | 5.24 |
| 7 | 0.17 | 4.08 | 6.50 | 1.10 | 3.58 | 0.36 | 1.23 | 1.18 | 2.47 | 4.75 | 4.62 | 1.73 |
| 8 | 5.29 | 4.60 | 3.36 | 0.79 | 3.64 | 1.81 | 0.22 | 4.13 | 0.25 | 4.74 | 5.87 | 3.97 |
| 9 | 4.38 | 2.25 | 0.25 | 0.00 | 2.53 | 3.29 | 1.99 | 3.76 | 1.14 | 0.31 | 3.71 | 7.78 |
| 10 | 2.24 | 1.91 | 1.00 | 1.31 | 2.12 | 2.11 | 1.74 | 2.91 | 1.32 | 1.36 | 0.14 | 5.84 |
| 11 | 2.03 | 1.15 | 3.58 | 1.20 | 1.80 | 1.20 | 2.51 | 1.56 | 0.65 | 2.84 | 4.32 | 3.57 |
| 12 | 4.38 | 3.57 | 3.11 | 0.51 | 0.57 | 1.35 | 0.48 | 1.83 | 0.03 | 3.50 | 7.58 | 9.12 |
| 13 | 2.66 | 4.47 | 5.61 | 0.21 | 3.49 | 1.88 | 1.00 | 2.51 | 1.51 | 3.34 | 6.58 | 6.00 |
| 14 | 0.09 | 1.99 | 3.94 | 1.26 | 3.08 | 3.04 | 1.39 | 2.14 | 0.05 | 0.64 | 0.12 | 2.51 |
| 15 | 2.91 | 1.45 | 2.65 | 0.25 | 1.90 | 1.33 | 1.64 | 2.07 | 0.00 | 0.02 | 1.79 | 0.61 |
| 16 | 2.65 | 0.73 | 0.99 | 0.85 | 0.44 | 0.86 | 0.10 | 1.49 | 0.56 | 0.44 | 1.39 | 2.03 |
| 17 | 0.35 | 0.42 | 0.62 | 0.79 | 0.13 | 0.48 | 0.50 | 0.23 | 0.43 | 0.43 | 0.97 | 5.81 |
| 18 | 0.54 | 0.05 | 1.05 | 0.25 | 0.12 | 0.54 | 0.40 | 1.38 | 0.96 | 0.75 | 3.15 | 5.27 |
| 19 | 0.45 | 1.12 | 0.04 | 2.26 | 0.19 | 0.86 | 0.33 | 0.61 | 2.68 | 2.65 | 2.17 | 1.85 |
| 20 | 2.16 | 0.42 | 0.57 | 4.05 | 0.77 | 1.19 | 1.49 | 1.19 | 1.14 | 1.36 | 0.37 | 1.85 |
| 21 | 1.75 | 1.26 | 0.47 | 2.18 | 1.48 | 1.53 | 0.66 | 0.54 | 0.41 | 0.19 | 0.97 | 2.03 |
| 22 | 0.37 | 1.40 | 1.22 | 0.08 | 1.12 | 0.43 | 0.23 | 0.44 | 0.42 | 1.04 | 0.87 | 0.56 |
| 23 | 0.75 | 0.32 | 0.46 | 2.44 | 0.17 | 1.83 | 1.96 | 0.81 | 1.64 | 0.64 | 1.30 | 2.72 |
| Ave. | 1.80 | 1.52 | 1.89 | 0.98 | 1.53 | 1.34 | 1.20 | 1.52 | 0.94 | 1.51 | 2.35 | 3.33 |

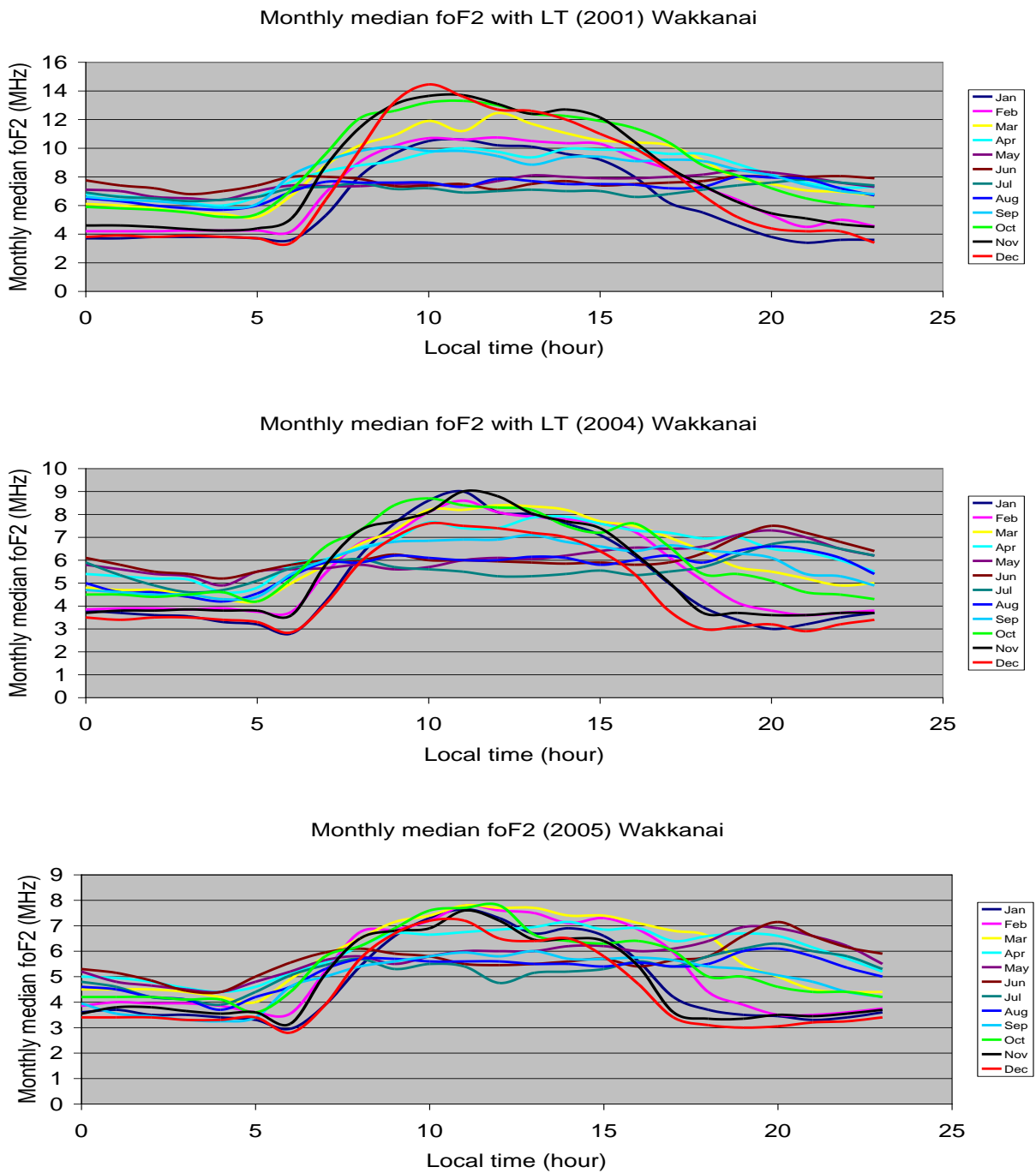

Figure 1. Observed Monthly median foF2 with local time for three years 2001, 2004, and 2005.

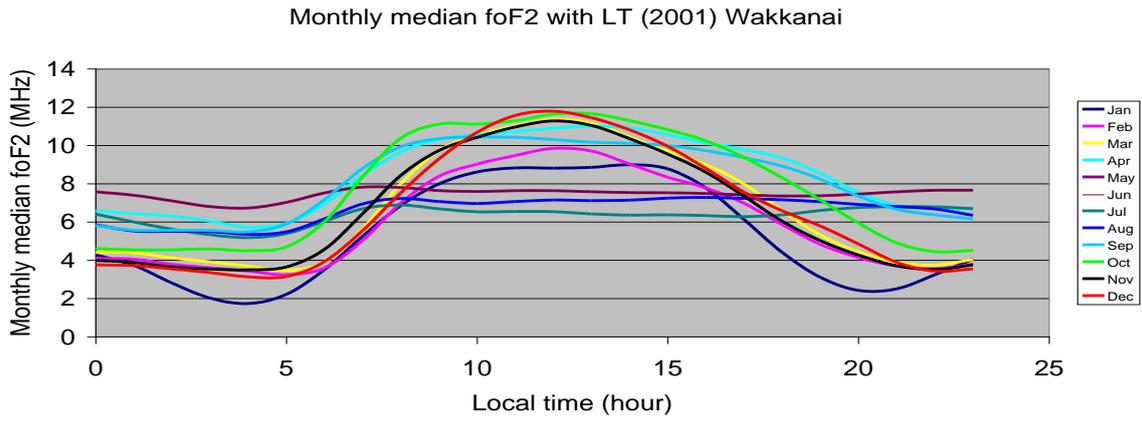

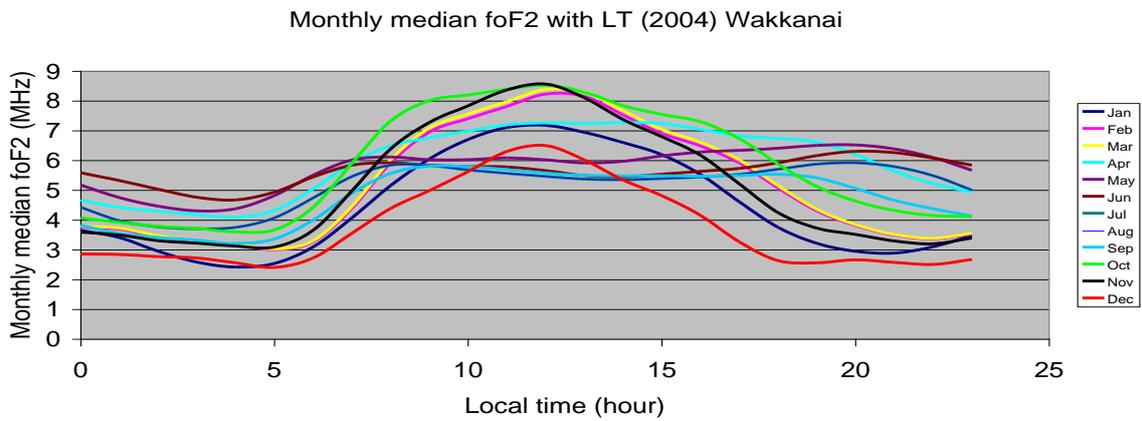

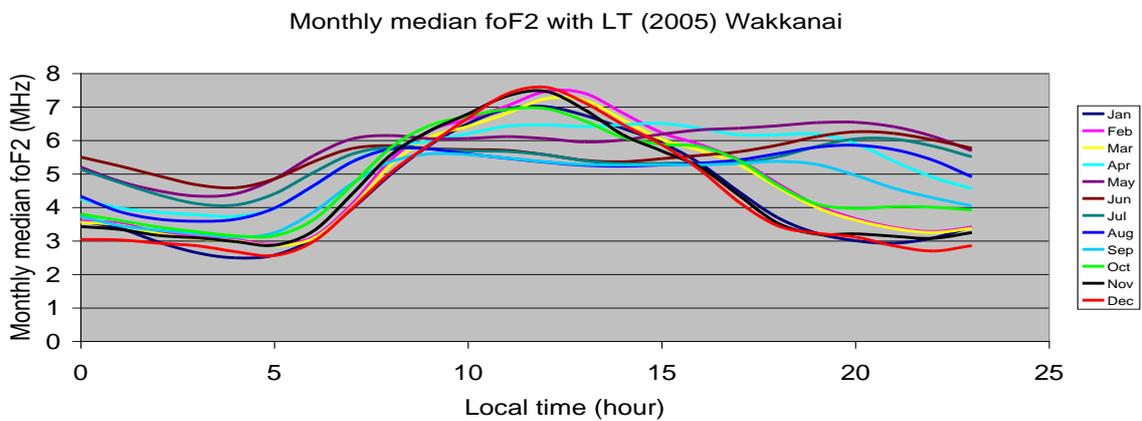

Figure 2. Predicted Monthly median foF2 with LT for years 2001, 2004, and 2005.

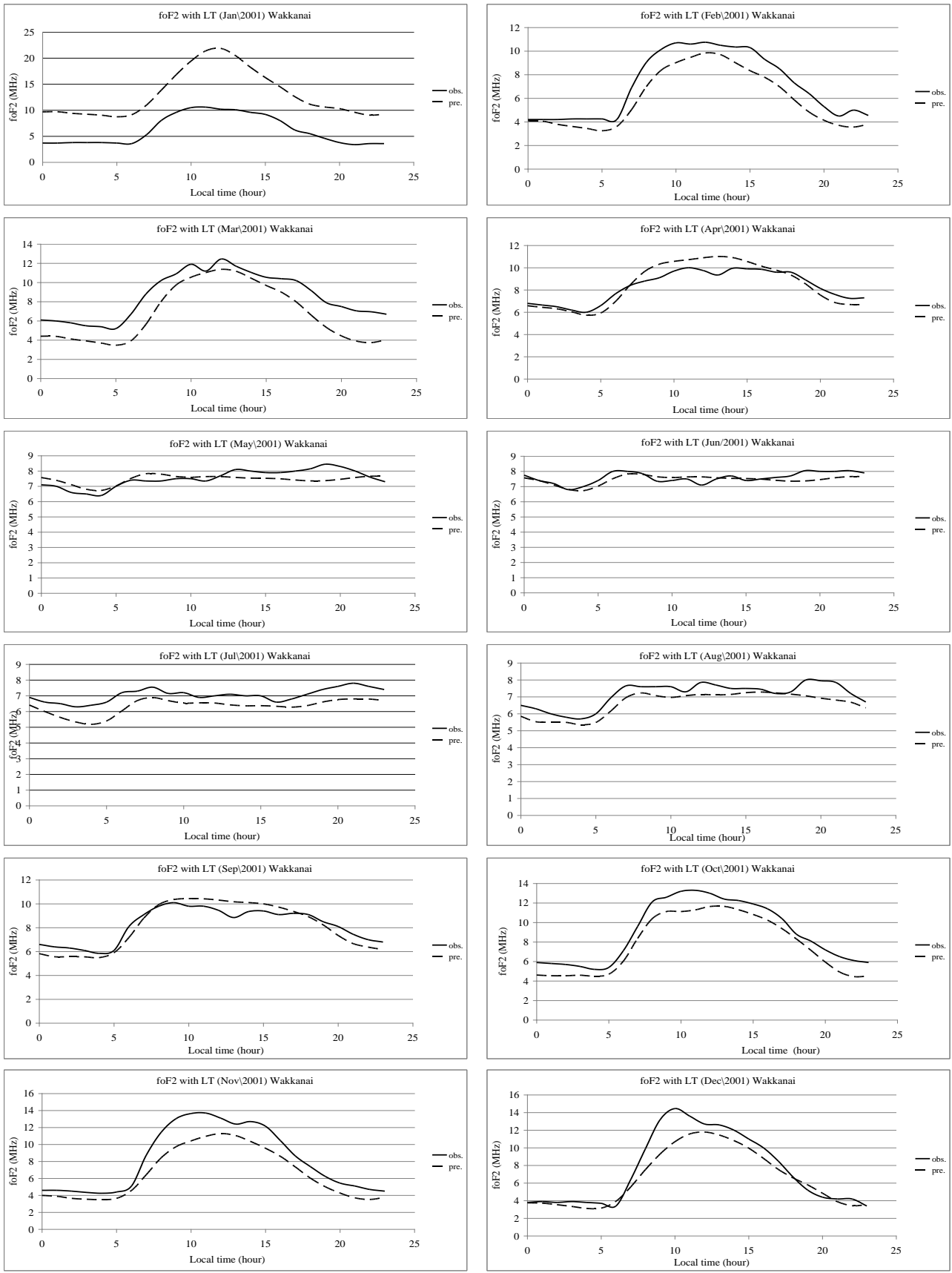

Figure 3. Observed and predicted foF2 with LT for year 2001.

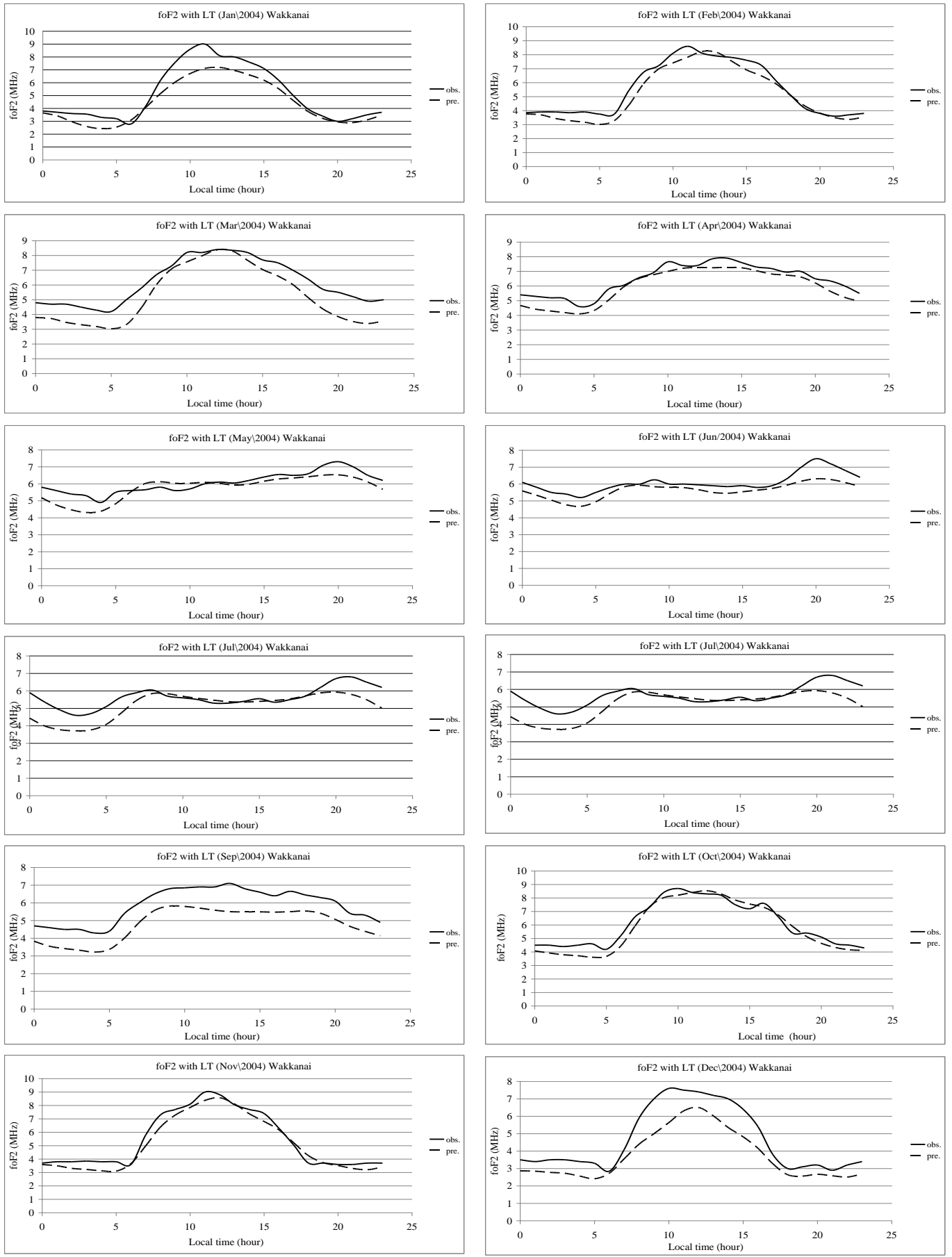

Figure 4. Observed and predicted foF2 with LT for year 2004.

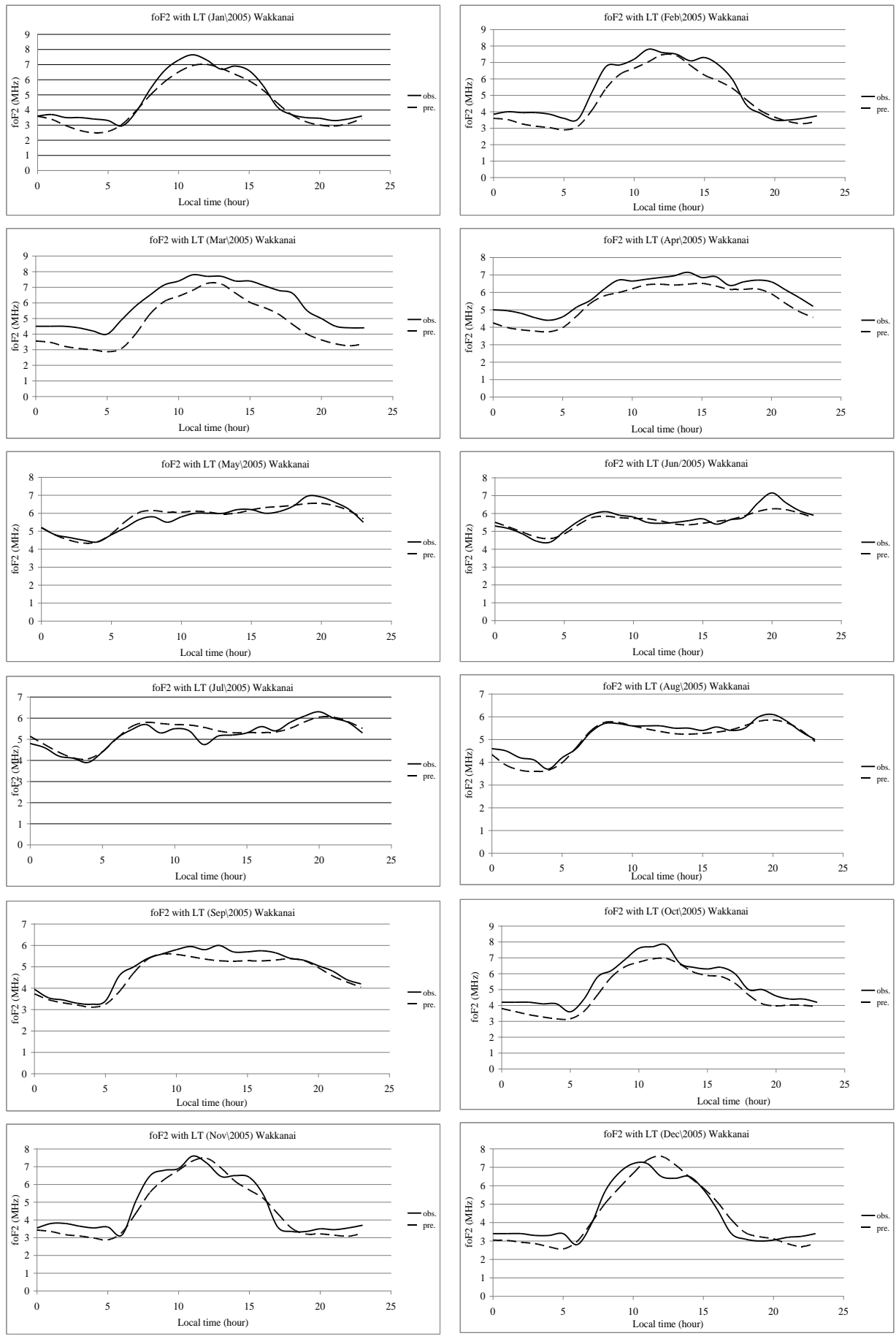

Figure 5. observed and predicted foF2 with LT for year 2005.

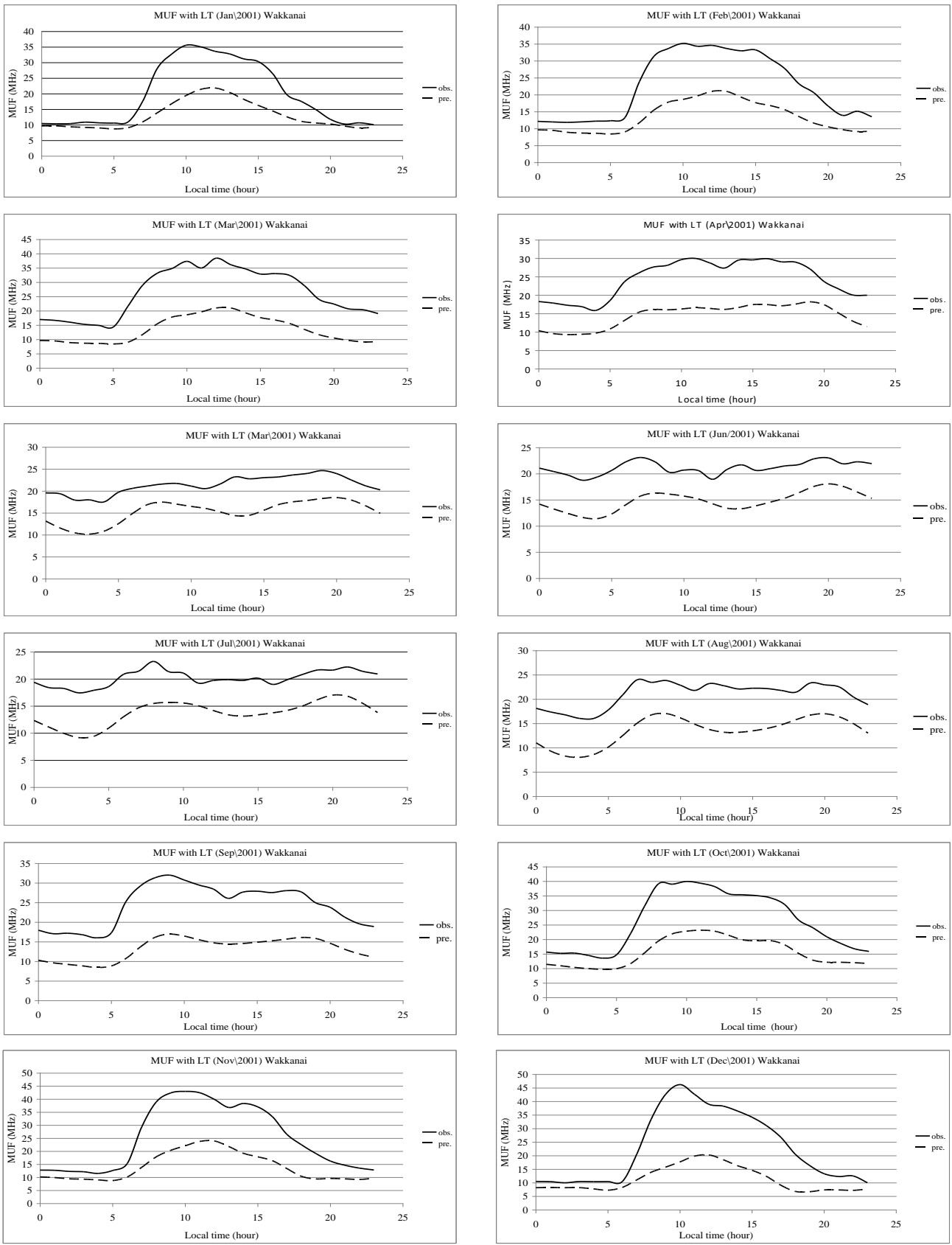

Figure 6 Observed and predicted BMUF with LT for year 2001.

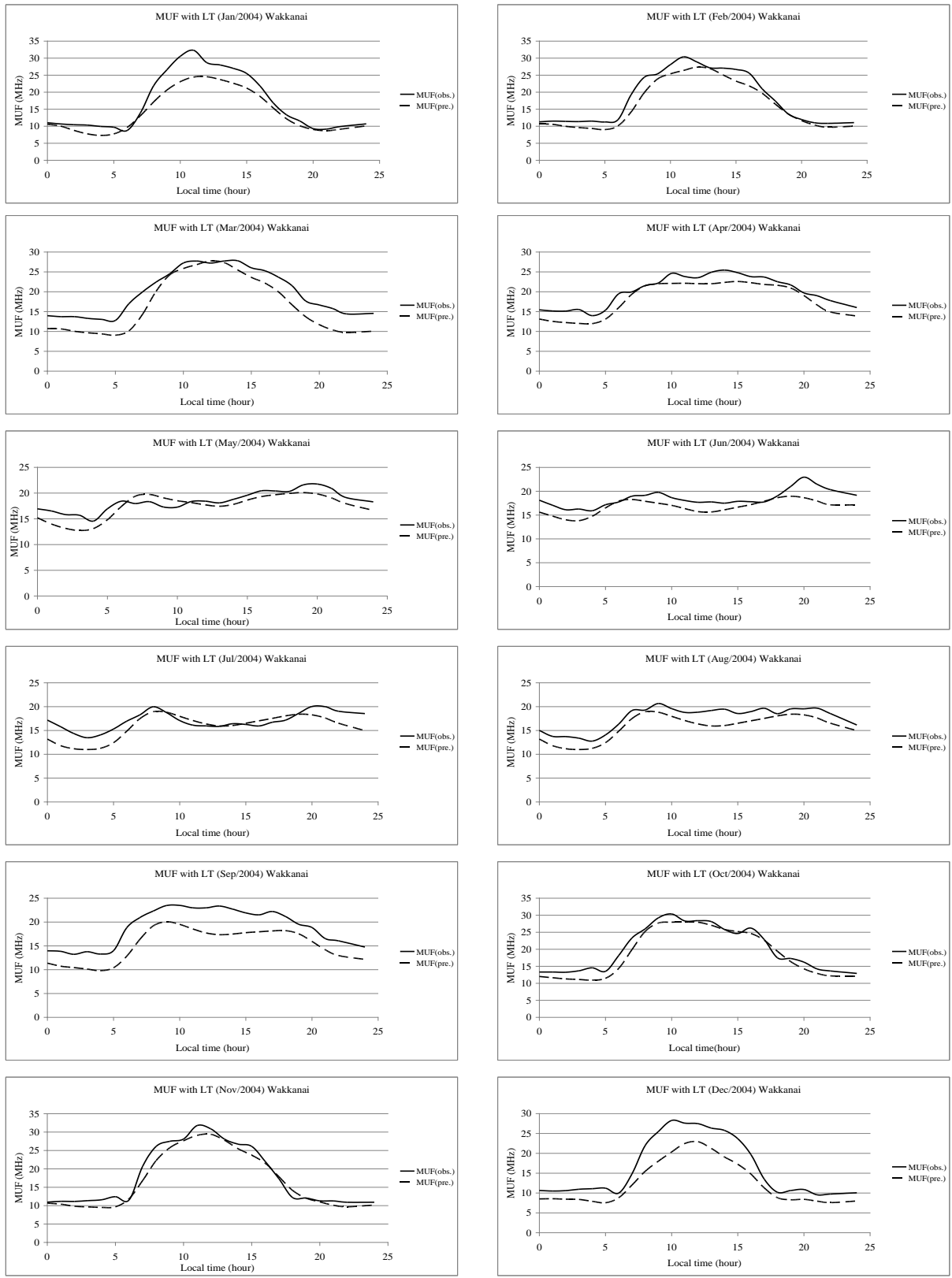

Figure 7 Observed and predicted BMUF for year 2004.

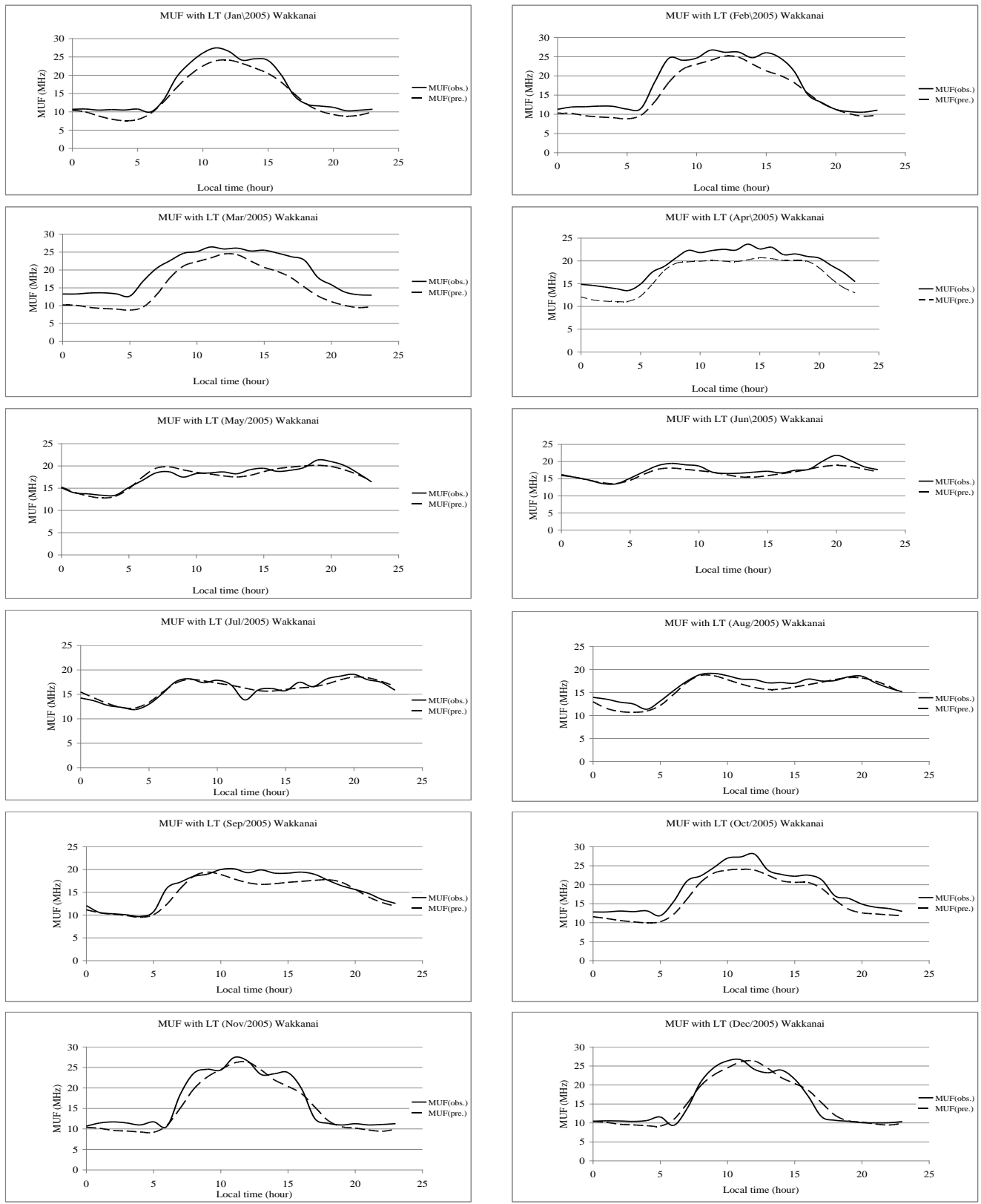

Figure 8 Observed and predicted BMUF for year 2005.

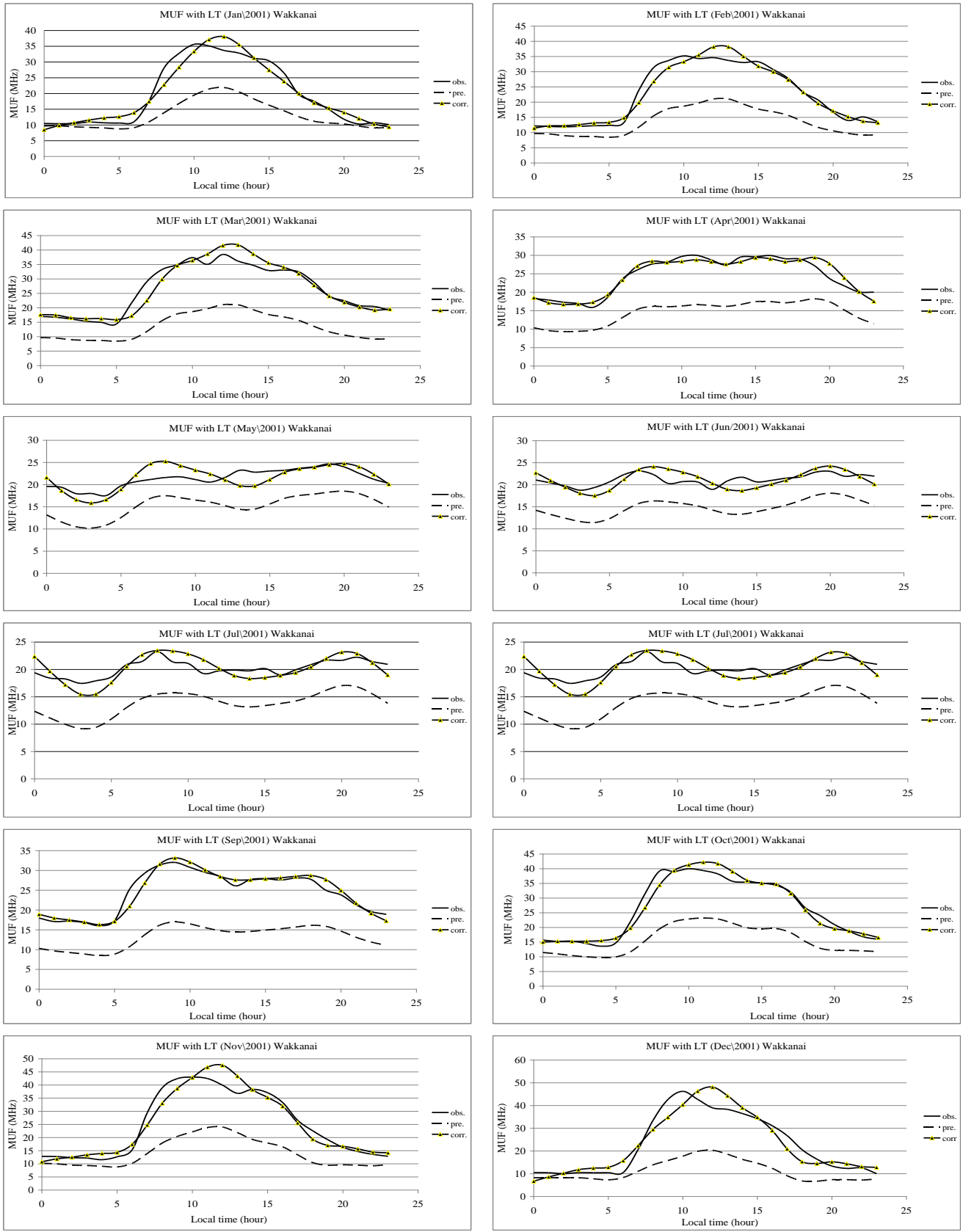

Figure 9 observed, predicted and corrected BMUF for year 2001.